Banner appropriate to article type will appear here in typeset article

# Hydrodynamics of an oscillating cylinder inline with steady current

**Chengjiao Ren[1], Feifei Tong[3,1], Fei He[1] and Liang Cheng[2, 1]**†

[1]School of Engineering, The University of Western Australia, 35 Stirling Hwy, Crawley, WA 6009, Australia

[2]School of Marine Science and Engineering, South China University of Technology, Guangzhou, 511442, China

[3]Faculty of Science and Engineering, Southern Cross University Gold Coast Campus, Gold Coast, QLD 4225, Australia

Wake and force characteristics of an oscillating cylinder in inline steady currents are investigated numerically over a wide parameter space of dimensionless oscillation amplitude ($A^* = 0.01 - 0.50$) and wavelength ($\lambda^* = 0.4 - 25$) at a fixed Reynolds number $Re = 500$. Fundamental issues addressed in this study are the interactions of wakes induced by steady approaching flow and cylinder oscillations and the influences of the governing parameters of $A^*$ and $\lambda^*$ on such interactions. Whilst the collinear flow is dominated by wakes induced by cylinder oscillation at $\lambda^* \lesssim 1.5$ and steady current at $\lambda^* \gtrsim 10$, it exhibits characteristics of nonlinear interactions of wakes induced by the cylinder oscillation and steady current at $\lambda^* \approx 1.5 - 10$, such as the formation of multiple synchronised modes interleaved with desynchronised modes. The synchronised mode varies with both $\lambda^*$ and $A^*$, forming an inclined Arnold's tongue across $\lambda^* - A^*$ space. There is a variety of the vortex shedding patterns in each synchronised mode. Variations of hydrodynamic force coefficients with $A^*$ and $\lambda^*$ are investigated with physical interpretations based on the wake characteristics. The applicability of Morison equation in predicting temporal variations of inline force is examined. We found that Morison equation shows reasonable accuracy only for a small range of $\lambda^* \lesssim 1.5$. Beyond this range, its performance deteriorates due to the influence of steady current on wake characteristics.

## 1. Introduction

Flow past an oscillating cylinder represents an idealised fluid-structure interaction problem in many engineering applications. For instance, an offshore water intake riser connected to a floating vessel is subjected to loading from the ocean current, where the riser may experience oscillations induced by the vessel motion. Another example is the submerged floating tunnel experiences combined loading from ocean current and oscillatory wave particles. A clear understanding of flow characteristics around and force responses on these cylindrical structures is of both fundamental and practical significance.

Above problem is often formulated as steady approaching current of velocity $U_\infty$ past an infinite long cylinder of diameter $D$ that undergoes forced oscillations with an oblique angle

† Email address for correspondence: liang.cheng@uwa.edu.au

**Abstract must not spill onto p.2**

| | | |
|---|---|---|
| Reynolds number | $Re$ | $U_\infty D/\nu$ |
| Oblique angle | $\theta$ | |
| Normalised amplitude | $A^*$ | $A/D$ |
| Normalised wavelength [1] | $\lambda^*$ | $\lambda/D = U_\infty/f_d D$ |
| Velocity ratio [1] | $r^*$ | $U_m/U_\infty = 2\pi A^*/\lambda^*$ |
| Frequency ratio [1] | $f^*$ | $f_d/f_0 = U_\infty/(\lambda^* f_0 D)$ |

TABLE 1. Non-dimensional parameters of flow past an oscillating cylinder. [1] The parameters of $\lambda^*$, $r^*$ and $f^*$ are interchangeable, one of which together with $A^*$ define the cylinder oscillation.

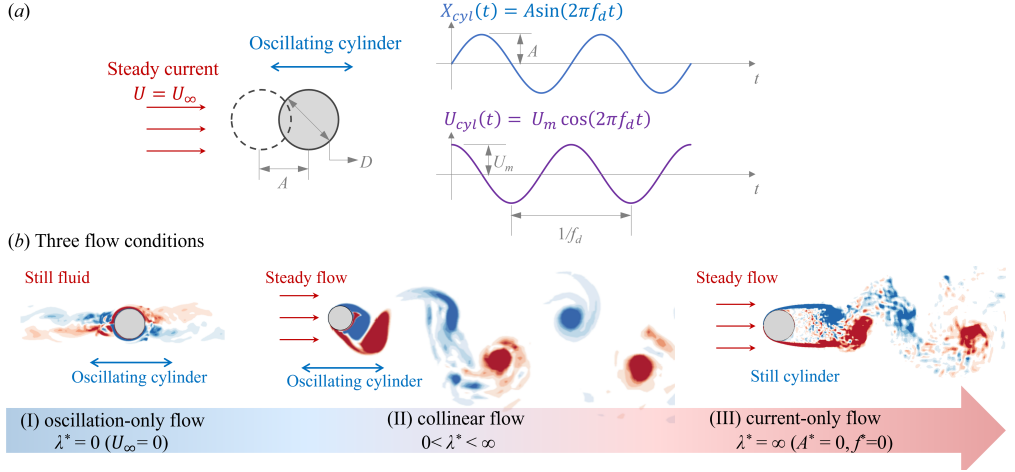

FIGURE 1. (a) Sketch of collinear flow, where $X_{cyl}(t)$ and $U_{cyl}(t)$ are the displacement and velocity of the oscillating cylinder, respectively. $A$ and $U_m$ are the amplitude of cylinder displacement and velocity, respectively. $f_d$ is the oscillation frequency. (b) Wake states of (I) oscillation-only condition at $\lambda^* = 0$, (II) collinear flow condition at $0 < \lambda^* < \infty$ and (III) current-only condition at $\lambda^* = \infty$.

$\theta$ to the direction of steady current. Cylinder oscillation is described by a sinusoidal function of time $t$ with amplitude $A$ and frequency $f_d$. Dimensional analysis reveals that the top four non-dimensional parameters listed in Table 1 can be chosen as the governing parameters of the problem. The $\lambda^*$ physically represents the normalised excursion distance of steady current in one cycle of cylinder oscillation. The velocity ratio $r^*$, or the frequency ratio $f^*$ can be used as an alternative parameter to $\lambda^*$, noting that $\lambda^*$, $r^*$ and $f^*$ are interchangeable together with $A^*$ (Table 1).

As the first step of an ongoing research effort, this paper considers a simple case of $\theta = 0°$ (termed as 'collinear flow'), as sketched in Fig. 1(a). Methods of investigation employed in previous studies on the topic are either experimental or numerical. While most of experimental studies were concerned with high $Re$ flows ($\sim O(1000 - 10,000)$) with narrow ranges of governing flow parameters, numerical studies were mainly carried out at low $Re$ ($<\sim 200$) through two-dimensional (2-D) direct numerical simulations (DNS) over a wide range of parameters. Differences between experimental and numerical results were found with regard to hydrodynamic forces and wake characteristics and will be discussed in following paragraphs. The causes for the observed differences in wake dynamics and hydrodynamic forces are unclear. The $Re$ effect and three-dimensionality of the flow are the likely culprits.

Research work published to date have reported the occurrence of multiple types of synchronised modes at intermediate $\lambda^*$. These modes are characterised by $p$ cycles of vortex shedding in $q$ cycles of cylinder oscillation, denoted as $p/q$ mode (Tang *et al.* 2017). Numerical work at low $Re$ reported the primary synchronisation mode (1/2 mode), featured by an anti-symmetric vortex shedding pattern repeating every two cylinder oscillation periods, dominated for $f^* \approx 2$ ($\lambda^* \approx 2.5$) (e.g., Leontini *et al.* 2011). Wakes with shedding of symmetric vortices (0/1 mode) and anti-symmetric vortices (1/1 mode) over each oscillation cycle develop when $\lambda^* \approx 2.0$ and $6.0$, respectively (Tang *et al.* 2017; Kim & Choi 2019). However, wake synchronisation modes reported in existing experimental studies show conflicting observations. Ongoren & Rockwell (1988) and Xu *et al.* (2006) observed the symmetric wakes (0/1 mode) at large $\lambda^*$ ($\approx 4.0 - 8.0$), in a region where 1/2 and 1/1 modes dominate at low $Re$. Though independent experiments reported the occurrence of 1/2 mode, it was restricted in a narrow $\lambda^*$ range (from 2.5 to 3.0), arguably smaller than that reported in Kim & Choi (2019) ($\lambda^* = 3.0 - 5.0$ at $Re = 200$).

In terms of hydrodynamic forces, a few studies have reported non-monotonic variations across different synchronised modes, including time-averaged and root-mean-squared force coefficients as well as drag and inertia coefficients from Morison equation (Morison *et al.* 1950), but these studies are mostly limited to low $Re$ under 2-D assumptions (Konstantinidis & Bouris 2017; Kim & Choi 2019). Conducting numerical simulations for high $Re$ flows or simultaneously measuring force and cylinder motion signals in experiments for low-amplitude cylinder oscillation pose significant challenges (Sarpkaya 2010), which is partially responsible for a lack of comprehensive understanding of variations of hydrodynamic forces distribute across $\lambda^* - A^*$ space.

Moreover, although we qualitatively understand the flow asymptotically approaches to the cylinder oscillation-only case (panel I in Fig. 1*b*) and steady-current-only case (panel III in Fig. 1*b*) as $\lambda^*$ approaches 0 and $\infty$ respectively, the exact parameter range where collinear flows can or cannot be approximated as oscillation-only or steady-flow-only conditions remains unclear.

This paper aims to address the issues outlined above. A total of 590 cases of 3-D DNS are conducted at a fixed $Re = 500$, spanning a wider parameter space of $\lambda^*$ from 0.4 to 25.0 and $A^*$ from 0.01 to 0.5 than that reported in previous studies. This oscillation amplitude range is relevant to wave-induced motion of cylindrical structures used in offshore engineering, such as tension leg platform and water intake risers of floating liquefied natural gas platforms (Chaplin 2000; Sarpkaya 2010).

The remainder of the paper is organised in the following manner. Numerical approach is offered in §2. Characteristics of wake and forces characteristics with respect to $\lambda^*$ and $A^*$ will be introduced in §3 and §4, respectively. Section 5 discusses the effect of Reynolds number. Finally, major conclusions will be drawn in §6.

## 2. Methodology

### 2.1. *Numerical method*

The governing equations for steady flow past an inline oscillating cylinder are the non-dimensional incompressible Navier-Stokes (N-S) equations:

$$\begin{aligned}\boldsymbol{\nabla} \cdot \boldsymbol{u} &= 0; \\ \partial \boldsymbol{u}/\partial t &= -(\boldsymbol{u} \cdot \boldsymbol{\nabla})\boldsymbol{u} - \boldsymbol{\nabla} p + Re^{-1}\nabla^2 \boldsymbol{u} + \boldsymbol{a},\end{aligned} \quad (2.1)$$

where $\boldsymbol{u} = (u, v, w)$ is the velocity vector in $x$-, $y$- and $z$-directions, $t$ is the time and $p$ is the pressure. $D$ and $U_\infty$ are used to normalise the above equations. The origin of a Cartesian coordinate system $(x, y)$ is placed at the centre of the cylinder. The harmonic cylinder motion $X_{cyl}(t) = A \sin(2\pi f_d t)$ is considered using a moving frame fixed on the cylinder, which is achieved by introducing an additional forcing term $\boldsymbol{a}$ to the momentum equations. This approach avoids the difficulty with mesh deformations and has been employed in many studies (e.g., Leontini *et al.* 2006; Gao *et al.* 2023; Xiong *et al.* 2020).

Eq. (2.1) is solved in a quasi-3-D manner using Nektar++ (Cantwell *et al.* 2015), where a spectral/*hp* element method is employed in the cross-sectional direction ($x - y$ plane); and a Fourier expansion method is used in the spanwise direction ($z$-direction). In present simulations, a modified Legendre basis with 5 modes ($N_p = 4$ polynomial order) is used in each *h*-type element. A second-order time integration method is employed, together with a velocity correction scheme in the Galerkin formula (Karniadakis *et al.* 1991; Blackburn & Sherwin 2004). The velocity in different spanwise locations can be expressed through a combination of $N_z$ Fourier modes. This approach offers advantages in terms of efficiency and facilitates the parallelisation of algorithms (Bolis 2013), and has been widely used and validated in our previous bluff-body flow studies (Ren *et al.* 2019, 2020, 2024; Liu *et al.* 2023, 2024; He *et al.* 2024).

A rectangular computational domain in a size of $(20D + 50D) \times (20D + 20D)$ is adopted. The steady current $U_\infty = 1$ is imposed in the *x*-direction. The free-stream velocity of $\boldsymbol{u} = (U_\infty, 0, 0)$ is specified in the inlet, top and bottom boundaries. The Neumann boundary condition ($\partial \boldsymbol{u}/\partial \boldsymbol{n} = \boldsymbol{0}$) is imposed in the outlet boundary. No-slip boundary condition ($\boldsymbol{u} = \boldsymbol{0}$) is enforced on the cylinder surface. A high-order Neumann pressure condition is specified on all domain boundaries, except that a reference zero Dirichlet pressure condition is employed on the outlet boundary. Zero initial conditions for velocities and pressure are employed in the simulations. In the spanwise $z$-direction, $L_z = 3D$ and $N_z = 64$ are used to capture the spanwise structure. Detailed mesh distributions and validations can be found in Ren *et al.* (2022) and are not shown here.

### 2.2. *Force coefficient definitions*

The total inline force coefficient ($C_x$) and transverse force coefficient ($C_y$) acting on the cylinder are defined as,

$$C_x = 2F_x/(\rho D U_\infty^2), \quad C_y = 2F_y/(\rho D U_\infty^2), \qquad (2.2)$$

where, $\rho$ is the fluid density, $F_x$ and $F_y$ are the inline and transverse forces per unit length. Their time-averaged quantities are termed as $C_{x.avg}$ and $C_{y.avg}$. The root-mean-square coefficients are defined as,

$$C_{x.rms} = \sqrt{\frac{1}{N}\sum_{i=1}^{N}(C_{x,i} - C_{x,avg})^2}, \quad C_{y.rms} = \sqrt{\frac{1}{N}\sum_{i=1}^{N}(C_{y,i} - C_{y,avg})^2}. \qquad (2.3)$$

where $N$ is the total number of sampling points.

In this study, variations of drag and inertia force coefficients over the $\lambda^*$ and $A^*$ are examined in detail. They are decomposed based on independent force-velocity model (Moe & Verley 1980), assuming the time-averaged and fluctuation components of inline force are induced by the steady current and cylinder oscillation, respectively. The 3-term Morison equation (Ren *et al.* 2021) is used in the velocity model to represent the fluctuation components. This is because it surpasses the conventional Morison equation in representing the inline force of an oscillating cylinder (Ren *et al.* 2023a; Dorogi 2022). The inline force $F_x(t)$ is represented by,

$$F_x(t) = \underbrace{\frac{1}{2}\rho D C_{x.avg} U_\infty^2}_{(i)} \underbrace{-\frac{1}{2}\rho D C_{d1} U_m^2 \cos(2\pi f_d t + \pi/4)}_{(ii)}$$
$$+ \underbrace{\frac{1}{2}\rho D C_{d2}(-U_{cyl}|-U_{cyl}|)}_{(iii)} + \underbrace{\frac{\pi}{4}\rho D^2 C_m \frac{\mathrm{d}(-U_{cyl})}{\mathrm{d}t}}_{(iv)}. \quad (2.4)$$

Here, $F_x(t)$ is composed of (i) the time-independent drag induced by the steady current (steady drag); the time-dependent (ii) linear drag and (iii) form drag and (iv) inertia forces associated with the oscillating cylinder. The procedure for acquiring the four force coefficients in Eq. 2.4 include:

(1) The $C_{x.avg}$ is obtained by taking the time-average of $C_x(t)$ from DNS case.

(2) According to Ren et al. (2021), (ii) is approximated by assuming $C_{d1}$ from the Stokes-Wang solution (Wang 1968) to the first-order as,

$$C_{d1} \approx 4\sqrt{2}\pi^{3/2} K^{-1} (\lambda^*/Re)^{1/2}, \quad (2.5)$$

where $K(=2\pi A^*)$ is the Keulegan-Carpenter number.

(3) Subtracting (i) and (ii) from $F_x(t)$, $C_{d2}$ and $C_m$ can be determined from the remaining force using least-square method.

Eq. 2.5 expresses that for a fixed $Re$, the $C_{d1}K$ is linearly proportional to $(\lambda^*)^{-1/2}$, regardless the oscillation amplitude, term as linear drag factor. In a similar manner, $C_{d2}K$ rather than $C_{d2}$ will be used in this study to represent the form drag factor of the oscillating cylinder.

## 3. Wake characteristics

### 3.1. *Regime map*

Figure 2 provides a complete distribution of collinear wake over the $\lambda^* - A^*$ space. To assist comparisons with previous studies who use $f^*$ or $K$ to interpret wake characteristics (e.g., Olinger & Sreenivasan 1988; Leontini et al. 2011, 2013), values of $f^*$ and $K$ are included as the secondary axes in Fig. 2. All the simulations are carried out for at least 20 oscillation periods and are continued for an extra 20 oscillation periods after the flow becomes fully developed to acquire data for statistical analysis.

In general, the collinear flow can be largely categorised into three groups with small, large and intermediate $\lambda^*$. At small and large $\lambda^*$, wakes are primarily governed by the cylinder oscillation ($\lambda^* \lesssim 1.5$), and the steady current ($\lambda^* \gtrsim 10.0$), respectively. At intermediate $\lambda^* \approx 1.5 - 10.0$, complex interactions leading to formations of multiple $p/q$ synchronised modes are observed. Detailed wakes characteristics within each flow group will be introduced in subsequent subsections.

### 3.2. *Oscillation dominant flows*

Characteristics of an oscillation-only case is introduced first. Fig. 3(*a*) shows the time-histories of $C_x$ and $C_y$ (left panel) and instantaneous spanwise-averaged flow field (right panel) for a case at $A^* = 0.40$ and $f^* = 4.0, (U_m = 2.08)$. Time histories of $C_x$ and $C_y$ in Fig. 3(*a*) are normalised by $U_\infty$ rather than the $U_m$ to allow a direct comparison with the collinear flow in Fig. 3(*b*). The $C_x(t)$ oscillates periodically in each oscillation cycle, while the $C_y(t)$ fluctuates randomly with small magnitude (less 1% of the $C_x$ magnitude). The small value of $C_y(t)$ is associated with the formation of a nearly symmetric wake (Fig. 3*a*). When the cylinder moves to the downstream, a symmetrical pair of shear layers emerges and propagates upstream on the cylinder surface. In the next half period when the cylinder moves upstream, another pair of shear layers with opposite signs forms. The shear layers from the preceding half period subsequently wrap up the recent generated shear layers. According to the nomenclature in Elston *et al.* (2006), such a process weakly holds a reflection symmetry $K_y$, i.e., $\omega_z(x, y, t) = -\omega_z(x, -y, t)$ and a spatio-temporal symmetry $H_1$, i.e., $\omega_z(x, y, t) = -\omega_z(-x, y, t + T/2)$. The 'weakly' symmetry is because the 3-D structure develops and undergoes complex interactions into turbulence (Sarpkaya 2010).

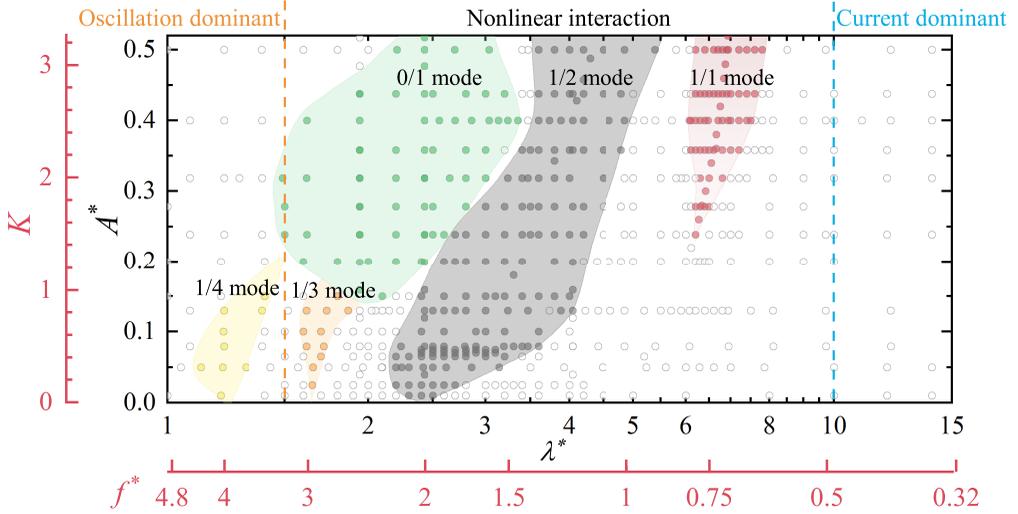

FIGURE 2. Distributions of the wake states over the space of $\lambda^* - A^*$ for $Re = 500$. The collinear flows are categorised intro three groups, namely the oscillation dominant flows at $\lambda^* \lesssim 1.5$, the nonlinear interaction flows at $1.5 \lesssim \lambda^* \lesssim 10.0$ and the current dominant flows at $\lambda^* \gtrsim 10.0$. Discrete circles are simulated cases.

When a steady current is introduced in the direction of cylinder oscillations, it modifies force and wake characteristics. An example is provided in Fig. 3(*b*) where the oscillation amplitude ($A^* = 0.40$) and frequency ($f^* = 4.0$) are the same as those in Fig. 3(*a*). Firstly, steady current convect the outer shear layers and shed vortices downstream, thereby disrupting the $H_1$ symmetry condition. Given the steady current velocity $U_\infty$ is relatively small compare to the $U_m$ ($\approx 2.07 U_\infty$), the near wake retains a $K_y$ symmetry. This is marked by the symmetric generation of shear layers on the cylinder surface, covered by long parallel shear layers extending until about $10D$ downstream before they interact to shed large coherent vortex structures (Fig. 3*b*). The near cylinder flow retains the feature of the oscillation-only wake. This is supported by time histories of $C_x(t)$ in the two scenarios (Fig. 3*a, b*), manifesting the similar magnitudes (only a 1.72% difference) and phase differences with $X_{cyl}(t)$.

The impact of steady current on the wake dynamics induced by the cylinder oscillation is further elaborated by considering the changes of pressure distributions around the cylinder surface and wake three-dimensionality with $\lambda^*$ for a fixed $A^* = 0.40$. First, the mean pressure $\overline{C_p}$ (normalised by $U_m$) around the cylinder surface between the oscillation dominant flows and the oscillation-only case are compared (Fig. 4). All the collinear flow cases in Fig. 4 reveal the same reflection symmetry along the $x$-axis ($\alpha = 180°$) as that of the oscillation-only case (the dashed line, $f^* = 4.00$). The symmetric distributions of $\overline{C_p}$ contours around the cylinder, as seen in the two insets at $f^* = 6.00$ and $3.00$, further support this observation. As $f^*$ decreases, corresponding to an increase in $\lambda^*$, the $\overline{C_p}$ deviates from the oscillation-only counterparts. Specifically, the $\overline{C_p}$ around the front stagnation point ($\alpha = 180°$) increases, while $\overline{C_p}$ around the back stagnation point ($\alpha = 0°$) is less affected. Additionally, the negative pressure on the side surfaces ($\alpha = 90°$ and $270°$) become larger as compared to that of the oscillation-only case.

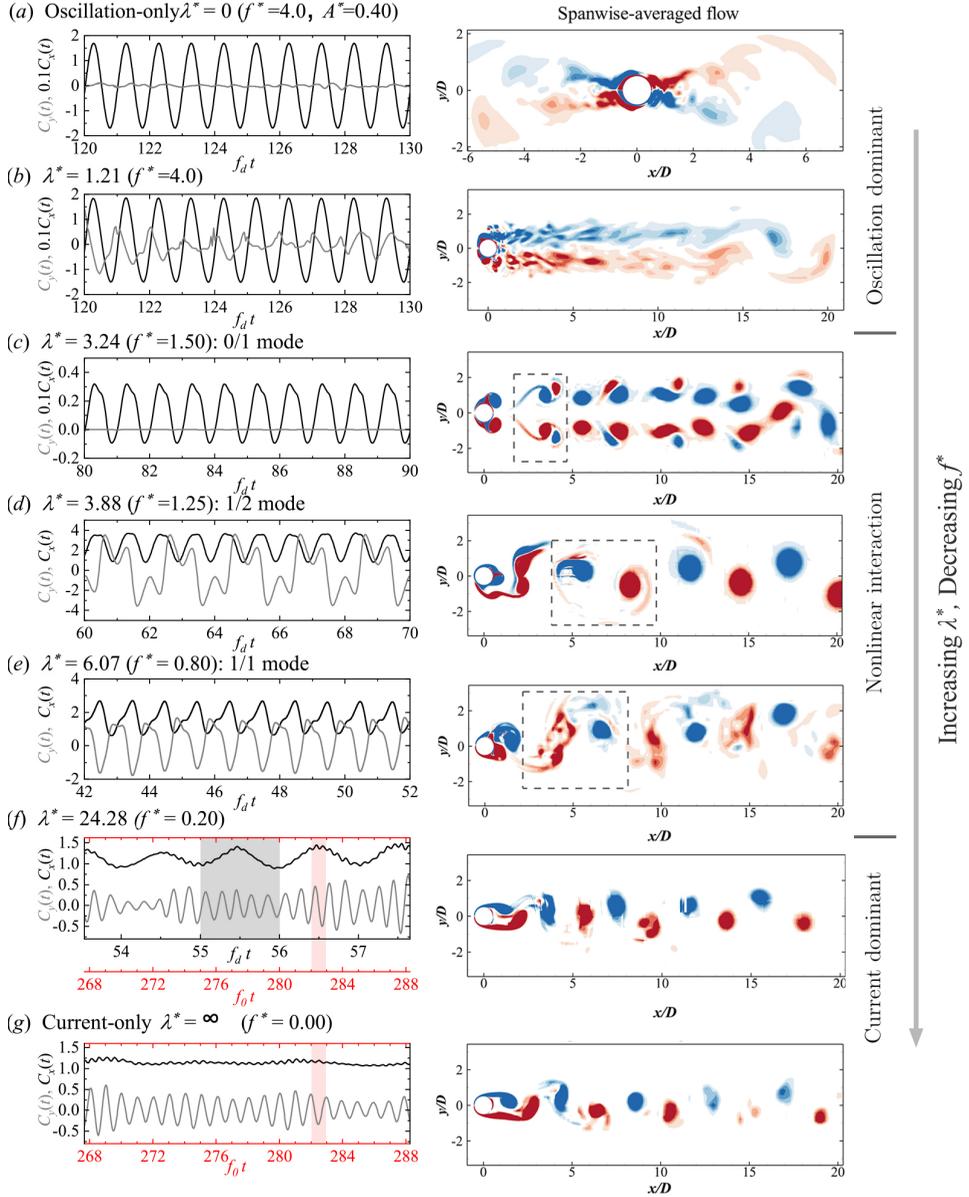

FIGURE 3. Force evolutions (left) and instantaneous wake patterns (right) of collinear flows at $A^* = 0.4$. The force coefficients are normalised by $D = 1$ and $U_\infty = 1$. The time units of the black horizontal axis in (a-f) are normalised by oscillation period ($1/f_d$), while time units of the red horizontal axis in (f-g) are normalised by the natural vortex shedding period ($1/f_0$). Cases in (a) and (g) are the corresponding oscillation-only ($U_\infty = 0$, $f^* = 4.0$) and current-only cases. The dashed grey boxes in the right panel of (c-e) indicate vortex formation in one wake synchronisation cycle.

Secondly, the steady current is found to increase the three-dimensionality of the wake induced by the cylinder oscillation. This is initially suggested by the finer flow structures appear in the spanwise-averaged flow (Fig. 3b) compared to the oscillation-only counterpart (Fig. 3a). To quantify this, the ratio of transverse enstrophy $\varepsilon_{xy}$ over total enstrophy $\varepsilon_t$ is used to represent the degree of near wake three-dimensionality between two scenarios. The enstrophies are defined as,

$$\varepsilon_{xy} = \frac{D^2}{2U_\infty^2 \Omega} \int_\Omega (\omega_x^2 + \omega_y^2) d\Omega, \quad \varepsilon_t = \frac{D^2}{2U_\infty^2 \Omega} \int_\Omega (\omega_x^2 + \omega_y^2 + \omega_z^2) d\Omega, \quad (3.1)$$

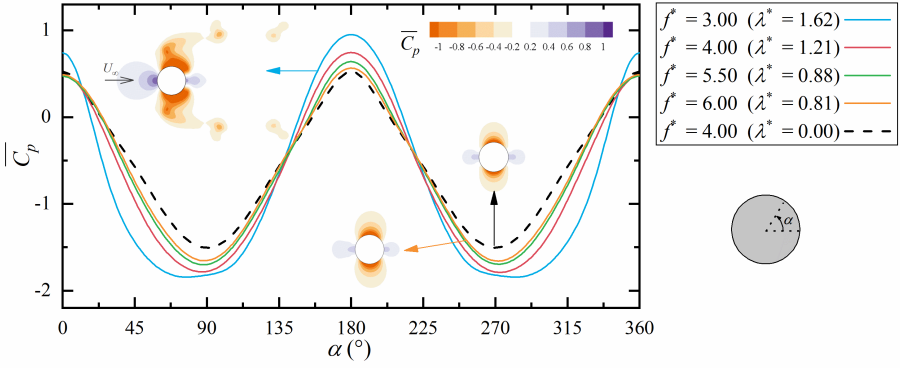

FIGURE 4. Mean pressure coefficient distributions $\overline{C_p}$ on the cylinder surface for collinear flows at $f^* = 3.00 - 6.00$ ($\lambda^* = 0.81 - 1.62$) and an oscillation-only case at $f^* = 4.0$. The horizontal axis corresponds to the circumferential angle $\alpha$ measured from the rear point of the cylinder. The dashed black line is oscillation-only case in Fig. 3(a). The $\overline{C_p}$ contours around the cylinder for the oscillation-only case and two collinear cases are included in the plot for comparison.

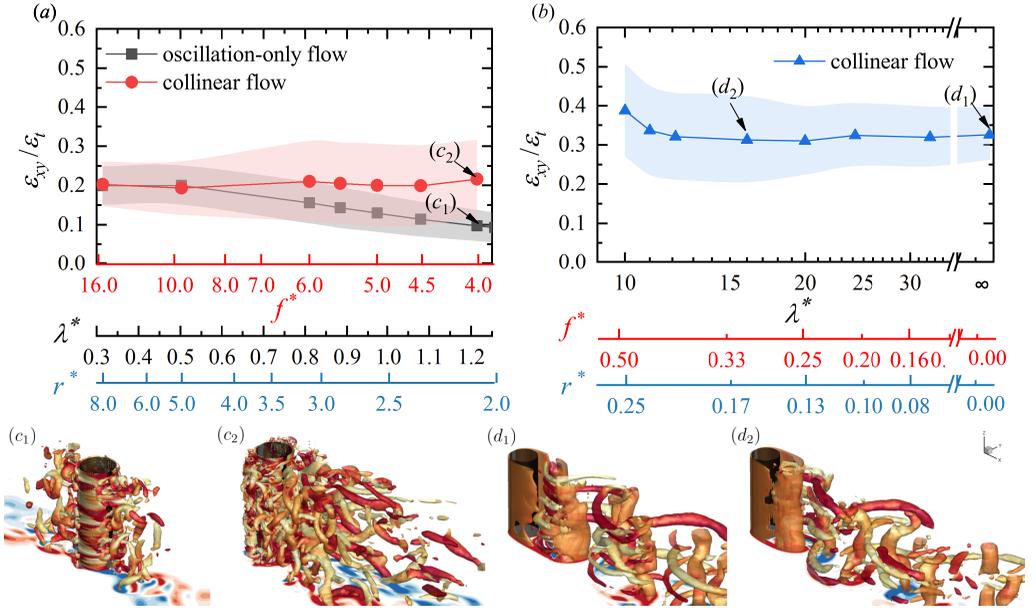

FIGURE 5. Distributions of $\varepsilon_{xy}/\varepsilon_t$ for (a) oscillation dominant flows at $f^* = 4.0 - 16.0$ and (b) current dominant flows at $\lambda^* = 9.71 - 32.4$. The solid lines represent the time-averaged quantity $\overline{\varepsilon_{xy}/\varepsilon_t}$ while the shaded area represents $\overline{\varepsilon_{xy}/\varepsilon_t} \pm \{\varepsilon_{xy}/\varepsilon_t\}'$. The oscillation-only flow counterparts at identical $f^*$ and $A^*$ as the collinear flows are provided in (a); and the current-only flow counterparts at $Re = 500$ are plotted at $\lambda^* = \infty$ in (b). Instantaneous 3-D flow fields for ($c_1$) the oscillation-only and ($c_2$) collinear flow cases at $f^* = 4.00$, and ($d_1$) the current-only and ($d_2$) the collinear flow cases at $\lambda^* = 16.19$ ($f^* = 0.30$). The 3-D structures are represented by the $\Lambda_2 = -2$ iso-surfaces coloured by $\omega_x \in [-2, 2]$.

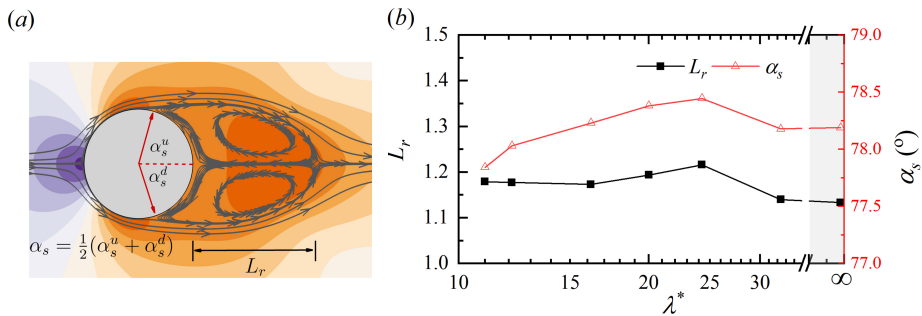

FIGURE 6. (*a*) Definitions of wake recirculation bubble length $L_r$ and averaged separation angle $\alpha_s$ using a time-averaged flow field at $\lambda^* = 32.38$ and $A^* = 0.40$. The streamlines are marked as dark grey lines with arrows. The purple and orange contours are the spanwise-averaged $\overline{C_p}$ field. (*b*) Distributions of $L_r$ (left axis) and $\alpha_s$ (right axis) for collinear flows at $\lambda^* \gtrsim 10.0$.

where the $\omega_x$, $\omega_y$ and $\omega_z$ are vorticity components in three directions; $\Omega$ is the integration volume at $x/D \in [-5, 5]$, $y/D \in [-5, 5]$ and $z/D \in [0, L_z]$, the region where the majority of coherent structures are distributed. The enstrophies were recorded every 20 non-dimensional time steps. A larger ratio indicates a stronger 3-D wake as $\varepsilon_{xy}/\varepsilon_t = 0$ for 2-D wake (Papaioannou *et al.* 2006; Ren *et al.* 2023*b*).

Fig. 5(*a*) details the mean and RMS of $\varepsilon_{xy}/\varepsilon_t$ for both collinear and oscillation-only flows at different $f^*$. At extremely large $f^*$, such as at $f^* = 15.50$, the steady current has a negligible effect on the 3-D wake induced by the cylinder oscillation, as evidenced by similar values for both scenarios. With reducing the $f^*$ for oscillation-only cases, both mean and RMS of $\varepsilon_{xy}/\varepsilon_t$ decrease, indicating that the instantaneous wake becomes less three-dimensional. This is because in the oscillation-only scenario, an increase in $f^*$ at a fixed $A^*$ corresponds to a decrease in the Stokes number $\beta$, causing the wake transition to two-dimensional (Sarpkaya 2010). In contrast, values of collinear flows at corresponding $f^*$ exhibit opposite trends, suggesting the enhancement of 3-D wake by the steady current. At the smallest $f^* = 4.0$ shown in Fig. 5(*a*), the mean and RMS $\varepsilon_{xy}/\varepsilon_t$ for the collinear case are nearly doubled. This increased wake three-dimensionality is also evidenced by the instantaneous 3-D fields for the collinear case in Fig. 5($c_2$) and oscillation-only case in Fig. 5($c_2$).

### 3.3. *Current dominant flows*

At large $\lambda^* \gtrsim 10.0$, a small cylinder motion slight modifies the classical Kármán wake, denoted as the current dominant flow. As a typical case shown in Fig. 3(*f*), the wake formation at $\lambda^* \approx 24.28$ resembles that of the flow past a fixed cylinder at $\lambda^* = \infty$ in Fig. 3(*g*). The $C_y(t)$ of two scenarios has similar fluctuation magnitude and frequency (see one fluctuation cycle of $C_y(t)$ in the red shaded area). In contrast, the cylinder oscillation significantly changes the $C_x(t)$. As shown by the black line in Fig. 3(*f*), the $C_x(t)$ forms large-amplitude of fluctuations over each oscillation period ($1/f_d$, grey shaded area) together with low-amplitude fluctuations occurring in every half of the vortex shedding period (red shaded area).

The impact of the perturbation induced by the cylinder motion on the Kármán wake is relatively weak at $\lambda^* \gtrsim 10$ and thus not easily identified from the instantaneous flow field. Two quantities from the time-averaged and spanwise-averaged flow field, namely the recirculation bubble length $L_r$ and the flow separation angle $\alpha_s$, are used to depict the changes. As defined in Fig. 6(*a*), the $L_r$ is measured from the tip of the cylinder, $(x/D, y/D) = (0.5, 0)$, to the stagnation point in the wake. The $\alpha_s$ is the average separation angle measured from the tip of the cylinder to the top and bottom separation points. Evolution of $L_r$ and $\alpha_s$ with $\lambda^*$ for $\lambda^* > 10.0$ are shown in Fig. 6(b). As $\lambda^*$ changes between 10 to $\infty$, the $L_r$ varies only slightly between 1.1 and 1.2 (left-$y$ axis); and $\alpha_s$ barely changes (within $1.0°$ in the right-$y$ axis). By introducing a small perturbation through oscillating the cylinder, such as reducing $\lambda^*$ from infinity to 24.28, the wake does become slightly wider (the increase of $\alpha_s$) and longer (the increase of $L_r$) compared with those of current-only scenario. With a further increase of the perturbation induced by the cylinder motion to a level where the nonlinear interactions become important (reducing the $\lambda^*$ to 10.0), both $L_r$ and $\alpha_s$ decline, indicating a narrower and shorter wake for smaller $\lambda^*$.

The influence of cylinder motion on the wake three-dimensionality is offered in Fig. 5(b) using $\varepsilon_{xy}/\varepsilon_t$. The mean $\varepsilon_{xy}/\varepsilon_t$ stays almost constant at around 0.33 as $\lambda^*$ decreases from $\infty$ to about 11.0, then it increases slightly to 0.4 as $\lambda^*$ further decreases to 10.0. However, the RMS value of $\varepsilon_{xy}/\varepsilon_t$ undergoes a gradual increase as $\lambda^*$ is reduced. This increase is because the vortex shedding is closer to the cylinder, resulting in stronger 3-D structures in the near wake region compared to the current-only case.

### 3.4. *Nonlinear interaction flows*

Nonlinear interactions between the vortex shedding induced by the current and the cylinder motion are pronounced at intermediate $\lambda^* \approx 1.5 - 10.0$, where multiple synchronised modes are identified as shaded areas in Fig. 2. The synchronised modes are highly dependent on the oscillation amplitude, occupying higher $\lambda^*$ ranges (smaller $f^*$) for higher $A^*$. The dominant $\lambda^* - A^*$ space of each synchronised mode resembles the Arnold's tongue in the nonlinear dynamical system (Pikovsky *et al.* 2003). Interleaved with those synchronised modes, desynchronised modes are developed (while circles in Fig. 2), whose vortex formations are aperiodic with the cylinder oscillation. Typical force and wake features for each synchronised mode will be discussed separately in the following paragraphs.

#### 3.4.1. *0/1 mode*

Fig. 3(c) shows a typical 0/1 mode at $\lambda^* = 3.24$ and $A^* = 0.40$, where shear layers on the top and bottom cylinder surfaces are symmetric about the oscillation axis from time to time. The vortices are shed symmetrically from the shear layers and are convected by the current to further downstream of the wake. As a result, the $C_y(t)$ in Fig. 3(c) is around zero; and the $C_x(t)$ fluctuates periodically over each oscillation cycle. The physics behind the development of 0/1 mode is that the Kármán wake is suppressed by the cylinder motion (Tang *et al.* 2017), i.e, the shear layers do not interact to form shed vortices in the near wake region. It was also referred to as 'symmetric mode' to indicate its symmetrical pattern in the near wake (Ongoren & Rockwell 1988; Zhou & Graham 2000).

In this study, 0/1 mode is defined as the wake retains symmetrical till $x/D \approx 5$ and $C_{y.rms} < 5 \times 10^{-4}$. In the regime map, the 0/1 mode is developed in oscillation dominant flows (Fig. 2) and occupies an area of $A^* \gtrsim 0.15$ and $\lambda^* \approx 1.5 - 3.3$. Increasing $A^*$ shifts the dominance region of 0/1 mode towards larger $\lambda^*$.

The vortex shedding patterns of 0/1 mode are elaborated using a group of cases at $\lambda^* \approx 1.94$ in Fig. 7 (a), where 0/1 mode develops at $A^* = 0.15 - 0.45$. Corresponding spanwise-averaged $\omega_z$ flow fields in Fig. 7(b) reveal that the number of shed vortices increases with $A^*$. The 0/1 mode at small $A^*$ is represented by the symmetric shedding of two single vortices per oscillation period, e.g., at $A^* = 0.159$ in Fig. 7($b_{II}$). This wake pattern is termed as S-I, following the naming convention in Xu *et al.* (2006). Increasing $A^*$ to $0.175 - 0.375$, one more vortex is generated from the cylinder surface alongside the main vortex, forming a pair of symmetric vortex pairs per oscillation cycle (Fig. 7$b_{II}$). To differentiate this vortex formation from S-I, it was termed as binary-vortex street or S-II in Xu *et al.* (2006). Further increasing the $A^*$ to 0.40, a new configuration is observed. Apart from the two vortex pairs symmetrically distributed on either sides of the $x$-axis, another row of counter-rotating vortices appears (see black dotted boxes in Fig. 7 $b_{IV}$ - $b_V$). Following the terminology used in Hu *et al.* (2019), it is termed as S-III herein. For $A^* > 0.477$, whilst the flow around cylinder surface is roughly symmetric, the downstream wake becomes chaotic (Fig. 7$b_{IX}$).

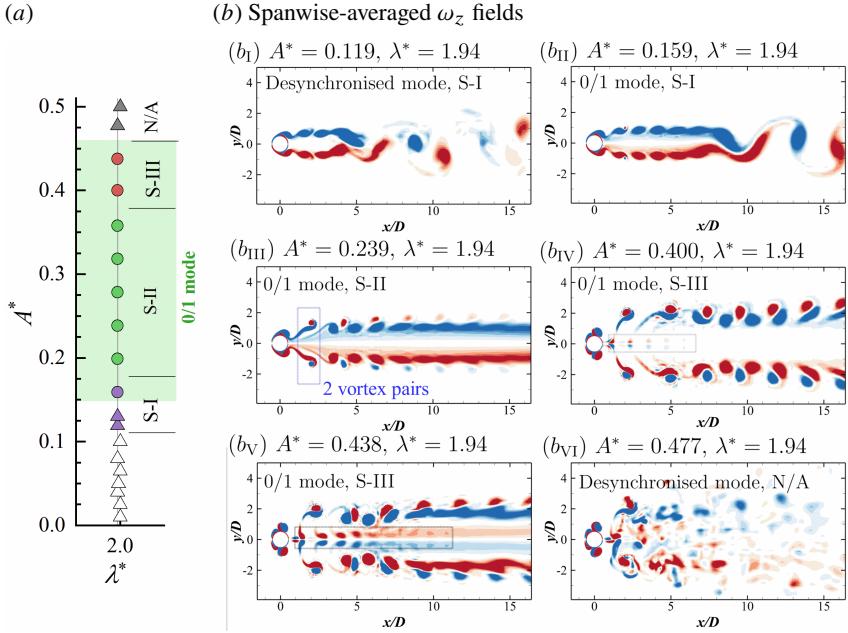

FIGURE 7. (a) Distribution of vortex formations for different $A^*$ at $\lambda^* = 1.94$. Symbols coloured by purple, green and red represent cases with S-I, S-II and S-III wake patterns, respectively. The dominant range of 0/1 mode, classified based on $C_{y,rms} < 5 \times 10^{-4}$, is shaded in light green. (b) Snapshots of the instantaneous vortex shedding patterns, visualised by the spanwise-averaged $\omega_z$ contours at levels between $\pm 1$.

The continuous increase of additional vortex pair(s) in each oscillation period as the wake transits from S-I via S-II to S-III is because the long oscillation period at large $A^*$ allows additional vortex pair(s) to generate on the cylinder surface. This phenomenon is similar to the increased number of shed vortices as $K$ increases in the oscillation-only scenario (Sumer & Fredsøe 1997; Tong *et al.* 2017).

3.4.2. *1/2 mode*

The 1/2 mode is characterised by periodic fluctuations of $C_y(t)$ in two periods of cylinder oscillation (refer to Fig. 3d). Hence, its vortex shedding pattern repeats every two cylinder oscillation cycles; and the FFT spectrum of $C_y(t)$ exhibits sharp peak on $f_y/f_d = 2$ (not shown). In the $\lambda^* - A^*$ space (Fig. 2), the 1/2 mode is initially found at $f^* \approx 2.0$ ($\lambda^* \approx 2.5$) for the smallest $A^*$ (= 0.01) examined herein, and displays an inclined distribution with increasing $A^*$, occupying a wider breadth for $f^* < 2.0$ than for $f^* > 2.0$. The envelop of 1/2 Arnold's tongue shifts entirely to $f^* \leqslant 2$ for $A^* \geqslant 0.2$. At highest $A^* = 0.5$ examined herein, the 1/2 mode dominates a range of $\lambda^* = 3.5 - 5.5$ ($f^* = 0.8 - 1.39$).

The vortex shedding pattern of 1/2 mode also reveals striking transition within the 1/2 Arnold's tongue. Given the strong dependency of 1/2 Arnold's tongue with $A^*$, vortex shedding formations along the centerline of 1/2 Arnold's tongue are presented in Fig. 8. At extremely small oscillation amplitude, such as at $A^* = 0.01$ in Fig. 8($b_I$), the 1/2 mode is featured by shedding of a single vortex on one side of the wake in one period and then another vortex with opposite sign on the other side of the wake in the next period. This wake formation maintains a spatio-temporal symmetry every two oscillation periods and is denoted as the 2S wake, following the naming convention by Williamson & Roshko (1988). As $A^*$ is increased to 0.05 in Fig. 8($b_\text{II}$), the 2S wake changes its formation: the braid shear layer connecting to major vortex core becomes stronger to form a secondary weak vortex each time the major vortex is shed. This 2S wake pattern is similar to the wake of an inline VIV at high reduced velocity (Fig. 9 in Zhao et al. 2022). Increasing $A^*$ to above 0.07, the 1/2 mode breaks the spatio-temporal symmetry to form a P+S wake (Fig. 8$b_\text{III}$). Similar asymmetric wake formation has been reported in Leontini et al. (2013); Tang et al. (2017) for low $Re$ flows. Tang et al. (2017) attributed it to the inherent wake asymmetry induced by the cylinder motion. In the case of an oscillating cylinder in still water, the lateral shedding of vortices to the axis of oscillation, such as Regime F and Regime D, have been well identified (Sumer & Fredsøe 1997). When $A^*$ is increased to around 0.20, one more vortex is shed from the cylinder surface. The wake is featured by a total of two vortex pairs every two oscillation periods and regains the spatio-temporal symmetry (2P wake, Fig. 8$b_\text{IV}$, $b_\text{V}$). The wake of 1/2 mode becomes more irregular for higher $A^*$(Fig. 8$b_\text{VI}$) and is termed as 2P* herein. The force evolution maintains the 2P wake character but the downstream wake is rather irregular and lack of periodicity.

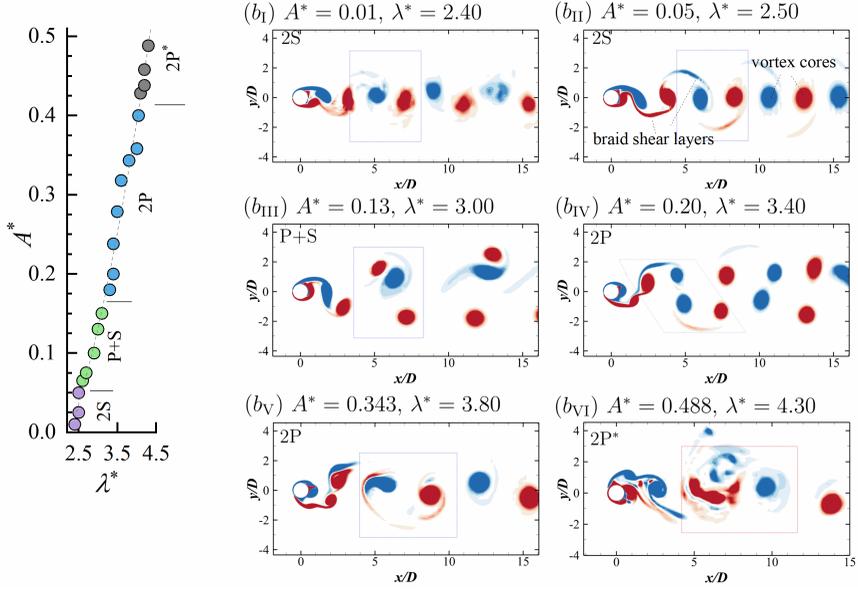

FIGURE 8. Similar to Fig. 7 but for distributions of vortex formations along the centerline of 1/2 Arnold's tongue (the dotted line in (a)). With increasing $A^*$, the vortex shedding of 1/2 mode transits in a sequence of 2S ($b_\text{I}$, $b_\text{II}$) → P+S ($b_\text{III}$) → 2P ($b_\text{IV}$ – $b_\text{V}$) → 2P* ($b_\text{VI}$). The dotted squares in (b) represent the vortex shedding patterns in every two periods.

### 3.4.3. *1/1 mode*

The 1/1 mode is featured by periodic fluctuations of both $C_x(t)$ and $C_y(t)$ within each cylinder oscillation cycle (Fig. 3e). The FFT spectra of both $C_x(t)$ and $C_y(t)$ manifest spark peaks at $f/f_d = 1$ (not shown). In Fig. 2, the 1/1 Arnold's tongue is centred around $f^* \approx 0.75$ and detected only for $A^* \geqslant 0.238$. The undetectable of 1/1 mode for $A^* < 0.238$ is possibly due to the narrower $\lambda^*$ range for smaller $A^*$, making it challenging to capture with the present resolutions of the $\lambda^* - A^*$ space.

The vortex formation of 1/1 mode is primarily governed by $\lambda^*$. Fig. 9 displays the instantaneous flow fields of 1/1 mode at $A^* = 0.4$. At the onset of 1/1 mode at $\lambda^* = 6.07$ (Fig. 9a), the wake is characterised by shedding of a pair vortices per oscillation cycle, termed as P wake. Precisely, each vortex is shed at the instant when the cylinder is travelling upstream or downstream. Because of the difference of combined velocity in each stroke, the two vortices formed in 1/1 mode are of unequal strength, resulting in a non-zero $C_{y.avg}$ and an biased wake pattern (slightly above $x$−axis in Fig. 9b). Increasing the $\lambda^*$ from 6.07 to 6.80, the wake becomes increasingly asymmetric to develop a P+S wake, featured by a secondary split of a strong vortex and its convection downstream on the other side of the cylinder (Fig. 9b). The $\Lambda_2$ iso-surface shown in Fig. 9 indicates that the P+S wake becomes more regular with weakened wake three-dimensionality until $\lambda^* = 6.80$. Increasing $\lambda^*$ further to 7.40, the three-dimensionality re-emerges and the wake asymmetry reduces (Fig. 9c). As the wake asymmetry develops gradually during the P → P+S transition, the boundary between P and P+S wakes is not delineated in Fig. 2.

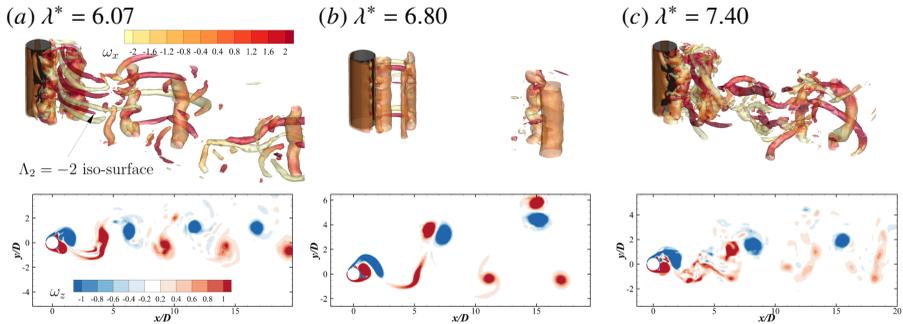

FIGURE 9. Instantaneous flows in the vicinity of 1/1 mode at fixed $A^* = 0.40$. In each subplot, the top panel shows the 3-D structure represented by the $\Lambda_2$ iso-surface coloured by $\omega_x$ contours; the bottom panel gives the spanwise-averaged $\omega_z$ field.

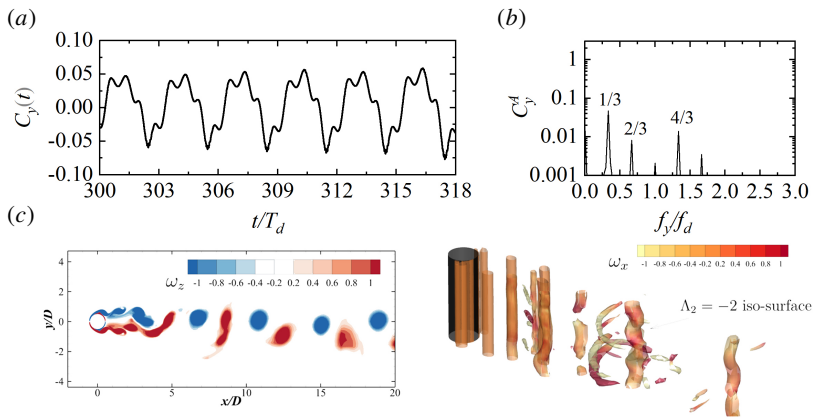

FIGURE 10. (a) Transverse force evolutions, (b) FFT spectra and (c) instantaneous flow field of a 1/3 mode at $A^* = 0.13$ and $\lambda^* = 1.619 (f^* = 3.0)$.

3.4.4. *Other synchronised modes*

Other modes, such as the 1/3 and 1/4 modes, are also identified in this study. These modes are observed at $A^* < 0.20$ when $f^*$ closes to 3 and 4, respectively. The occurrence of these modes is associated with the interaction of two incommensurable frequencies. For instance, two frequencies for the 1/3 mode are the forcing frequency $f_d = 3f_0$, and vortex shedding frequency $f_0$. Therefore, the $C_y$ FFT spectrum of a typical 1/3 mode(Fig. 10 b) reveals a sharp peak at the rational number of $f_y/f_d = 1/3$ and two double peaks near $f_y/f_d = 1$, resulting from frequency modulations. The wake of 1/3 mode resembles the S-I wake introduced before, characterised by symmetric parallel shear layers in the near wake region till $x/D \lesssim 2.5$ before interacting to form shed vortices in the downstream region.

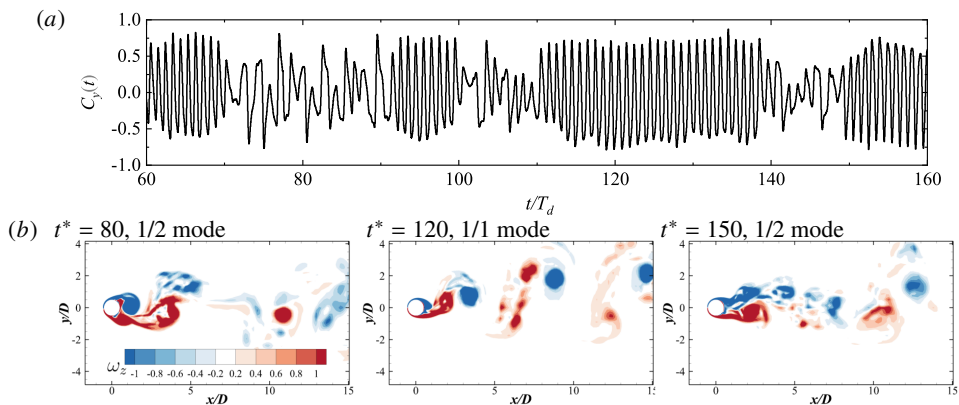

Figure 11. (*a*) Transverse force evolution and (*b*) instantaneous spanwise-averaged flow fields at $A^* = 0.318$ and $\lambda^* = 6.20$.

The 1/3 and 1/4 modes vaguely form Arnold's tongue shapes in Fig. 2. Other synchronised modes with relatively larger natural numbers of $p$ and $q$ (Tang *et al.* 2017) are not labelled in Fig. 2. This does not necessarily mean that they do not develop in the present $Re$, instead they require finer resolutions on parameters like $\lambda^*$ and $A^*$ to be captured than those currently employed. Moreover, it becomes extremely challenging when the wake becomes three-dimensional as the presence of spanwise dislocations can easily disrupt the high-order synchronisations. Given that $p/q$ modes with large integers occupy narrow parameter ranges with small impact on forces, see Fig.13 and Fig.14 in Tang *et al.* (2017), no further effort was made to detect all of them.

3.4.5. *Desynchronised mode*

An example of the desynchronised mode, located in the transitional region between 1/2 and 1/1 modes, is illustrated in Fig. 11. Both the flow fields and $C_y(t)$ display an intermittent swapping of 1/1 and 1/2 modes, termed as 'mode-switching' in Ongoren & Rockwell (1988).

Moreover, the development of 3-D structure favours the desynchronisation phenomenon. Similar to the effect of spanwise dislocations on the Kármán vortex street, the 3-D wake causes an irregular evolution of forces. Consequently, shed vortices do not perfectly synchronise with the cylinder's motion. Fig. 12 compares two cases at $A^* = 0.40$ and $\lambda^* = 5.71$ in both 2-D and 3-D simulations. While the 2-D flow in Fig. 12(*a*) exhibits a perfect 1/1 mode, its 3-D counterpart in Fig. 12(*b*) displays a more random evolution of forces.

4. **Hydrodynamic force coefficients**

Detailed distributions of hydrodynamic force coefficients with $\lambda^*$ and $A^*$ are illustrated. These statistical force coefficients include the time-averaged and RMS inline force coefficients ($C_{x.avg}$, $C_{x.rms}$), the time-averaged and RMS transverse force coefficients ($|C_{y.avg}|$, $C_{y.rms}$), the form drag factor ($C_{d2}K$) and inertia force coefficient $C_m$. Variations of the six force coefficients are primarily discussed using line plots at three $A^*$ groups through Fig. 13. Moreover, the contour plot of $|C_{y.avg}|$ generated by mapping results from 590 cases on the plane of $A^* = 0.01 - 0.5$ and $\lambda^* = 1.0 - 10.0$ is specifically given in Fig. 14 to indicate the transition of 2S → P+S → 2P wakes in the 1/2 Arnold's tongue. Contour plots for other force coefficients are offered in the appendix.

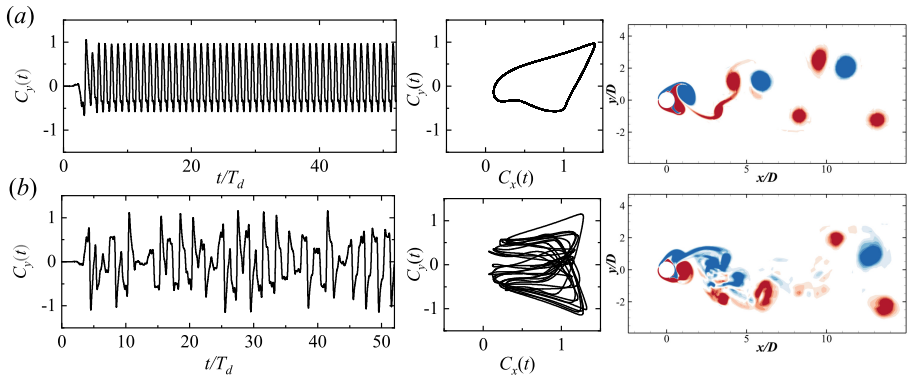

FIGURE 12. Instantaneous flow features of the flow at $\lambda^* = 5.71$ and $A^* = 0.40$ with (a) 2-D and (b) 3-D assumptions. In each subplot, the left panel represents the $C_y(t)$ time histories; the middle panel is the Lissajous diagram of $C_x(t) - C_y(t)$; the last panel shows a spanwise-averaged $\omega_z$ field at a time instant.

### 4.1. *Oscillation dominant flows*

A notable feature of collinear flows in the oscillation dominant group is that the inline force fluctuation components closely resemble those of oscillation-only flow. This feature is evident from the behaviour of $C_{x.rms}$, $C_{d2}K$ and $C_m$ in Fig. 13(*b, e, f*), where the curves of collinear flow at $A^* = 0.40$ (represented by black lines with symbols) collapse to oscillation-only counterparts at corresponding $f^*$ and $A^*$ (indicated by dashed grey lines). In the case of an oscillation-only scenario, Ren *et al.* (2021) demonstrated that the inline force primarily comprises of the linear drag force and inertia force for $K \lesssim 0.8$ and thus can be represented by the Stokes-Wang solution. As $A^*$ increases, the Stokes-Wang solution exhibits inaccuracies to the inline force on the cylinder in oscillation-only condition owing to the increased contribution of form drag to the total inline force (Ren *et al.* 2021). For the collinear flows at $\lambda^* < 1.50$, above demonstrations remain valid as suggested by the $C_{d2}K \approx 0$ and $C_m \approx 1$ at small $A^* < 0.13$ in Fig. 13(*e*) and (*f*), respectively. As $A^*$ increases, a slight increase of $C_{d2}K$ and a decrease of $C_m$ are observed, suggesting that the Stokes-Wang solution no longer represents the inline force of an oscillating cylinder. In the log-log plot of $C_{x.rms}$ (Fig. 13*b*), it linearly reduces with increasing $\lambda^*$; and a higher $A^*$ results in a larger $C_{x.rms}$ due to the increased contribution of $U_m$. The $|C_{y.avg}|$ (Fig. 13*c*) are nearly zero due to the symmetric generation of flow around the cylinder.

The $C_{x.avg}$ and $C_{y.rms}$, which are negligible in oscillation-only scenario, exhibit distinct variations due to the interference of steady current with the wake of oscillating cylinder. The $C_{x.avg}$ in Fig. 13(*a*) shows a continuous increase with increasing $\lambda^*$, which aligns with the $\overline{C_p}$ distribution in Fig. 4. Affected by complex vortex splitting and merging processes in the wake, the $C_{y.rms}$ varies drastically with both $\lambda^*$ and $A^*$.

### 4.2. *Current dominant flows*

In the current dominant flows ($\lambda^* \gtrsim 10$), the cylinder oscillation has minor effect on $C_{x.avg}$, $|C_{y.avg}|$ and $C_{y.rms}$. They closely resemble the current-only counterparts regardless of $\lambda^*$ and $A^*$ (see the grey horizontal line at $\lambda^* = \infty$ in Fig. 13). This is because the frequency of cylinder oscillation within this $\lambda^*$ range remains small compared to that of the Kármán wake ($f^* < 0.5$); and the Kármán wake is less affected by the oscillating amplitude. However, the $C_{x.rms}$ is amplified by the increase in $A^*$ and is larger than both current-only and oscillation-only counterparts.

The decomposed drag and inertia coefficients from Morison formula in Fig. 13(*e, f*) display almost monotonic variations with $\lambda^*$. An increase of $\lambda^*$ tends to increase $C_{d2}K$ and decrease $C_m$. Notably, the cylinder oscillation velocity is a small portion compared to that of the $U_\infty$;

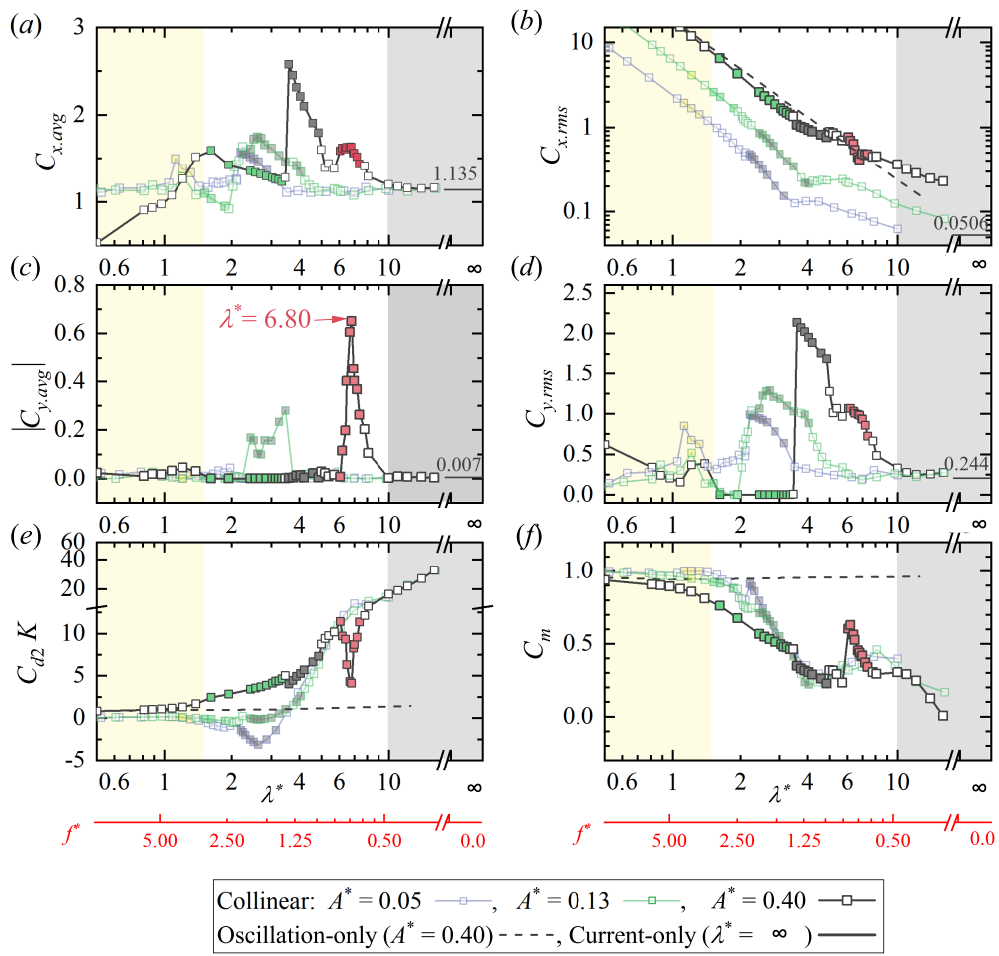

FIGURE 13. Distributions of (a) time-averaged inline force coefficient $C_{x.avg}$, (b) RMS inline force coefficient $C_{x.rms}$, (c) time-averaged transverse force coefficient $|C_{y.avg}|$, (d) RMS transverse force coefficients $C_{y.rms}$, (e) form drag coefficient $C_{d2}K$ and (f) inertia coefficient $C_m$, as functions of $\lambda^*$ for three different $A^*$. Corresponding values of the frequency ratio $f^*$ are plotted as additional horizontal axes below subplots (e) and (f). The grey solid lines at $\lambda^* = \infty$ indicate the corresponding values for the current-only case at $Re = 500$. The dashed grey lines in (b), (e) and (f) are the results of oscillation-only cases at corresponding $f^*$ for $A^* = 0.40$. Symbols filled with green, grey and red colours indicate occurrence of 0/1, 1/2 and 1/1 modes, respectively.

and the inline force evolution is affected by both $f_d$ and $f_0$, contradicting with the assumption of Morison equation. We refrain from placing significant emphasis on explaining the physics behind this, though similar variations were reported in Konstantinidis & Bouris (2017) at $A^* = 0.1$ and $Re = 150$.

### 4.3. Nonlinear interaction flows

In the nonlinear interaction flows ($1.5 < \lambda^* < 10$), force coefficients vary considerably with $\lambda^*$ and $A^*$ due to formation of multiple synchronised modes and vortex shedding patterns. In particular:

• In the 0/1 Arnold's tongue, the value of $C_{y.rms}$ is negligible ($< 5 \times 10^{-4}$) due to the symmetric creation of shear layers around the cylinder. The $C_{x.rms}$ departs from the oscillation-only counterpart as $\lambda^*$ increases beyond 1.50 (see the curves at $A^* = 0.40$ in Fig.

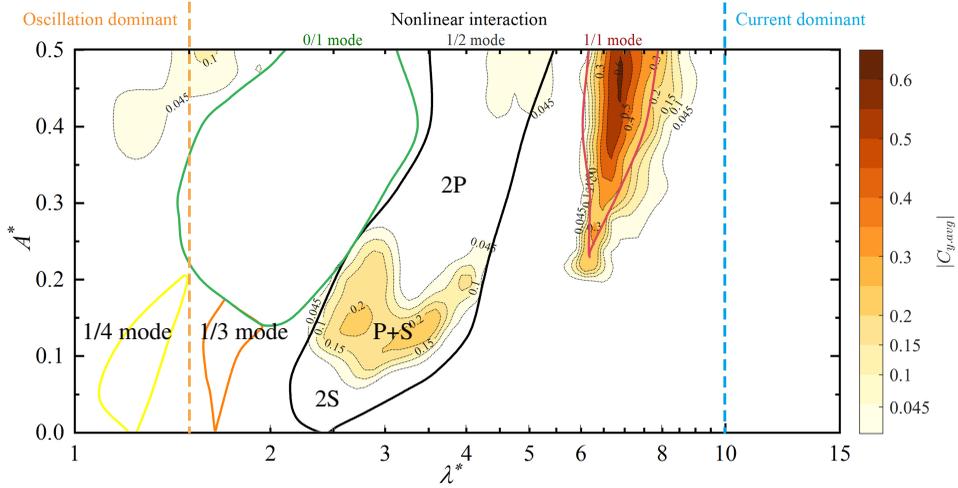

FIGURE 14. Contour of $|C_{y.avg}|$ over the $\lambda^* - A^*$ space. Boundaries of Arnold's tongues are indicated by solid lines.

13$b$), indicative of the modification of steady current to the wake induced by the cylinder oscillation. The continuous rise in $C_{d2}$ and decrease in $C_m$ are observed in Fig. 13($e$) and ($f$), respectively. At the end of 0/1 mode for $A^* = 0.4$ (around $\lambda^* = 3.35$), the $C_{d2}K$ reaches approximately five times the value of the oscillation-only case and the $C_m$ reduces to roughly half of the oscillation-only case.

• In the 1/2 Arnold's tongue, the $C_{x.avg}$ and $C_{y.rms}$ (Fig.13$a,d$) exhibit sudden surge at the lower $\lambda^*$ bound, and form a monotonic decrease trend with increasing $\lambda^*$. The surge of $C_{x.avg}$ and $C_{y.rms}$ at the onset of 1/2 mode is related to the enhancement of shear layers, forming a more organised wake flow compared with desynchronised modes (Tang *et al.* 2017). Hence, the value of $C_{y.rms}$ is enhanced greatly. Because the 1/2 Arnold's tongue is inclined over the $\lambda^* - A^*$ space, the crests of these two coefficients shift to larger $\lambda^*$ for higher $A^*$. The values of $C_{x.rms}, C_{d2}K$ and $C_m$ show slight initial reductions. After that, their variations with respect to $\lambda^*$ generally align with those of the 0/1 mode. By the end of the 1/2 mode, $C_{d2}K$ reaches to around 7.29, while $C_m$ lowers to around 0.25. The formation of P+S wake results in a large $|C_{y.avg}|$, such as at $\lambda^* \approx 2.5 - 4.0$ in $A^* = 0.130$ (the green curve in Fig.13$c$). The boundary of the P+S wake over the $\lambda^* - A^*$ space can be broadly indicated by the intensified contour of $|C_{y.avg}| \geqslant 0.045$ in the middle of the 1/2 Arnold's tongue (Fig. 14). Below and above this area indicate the dominant regions for 2S and 2P wakes.

• The 1/1 mode displays local-amplified $C_{x.avg}$ and $C_{y.rms}$ due to the synchronisation behaviour (Fig. 13$a, d$). In addition, $|C_{y.avg}|$ increases greatly within the 1/1 dominated region because of the occurrence of asymmetric wake. The intensified $|C_{y.avg}|$ contour in Fig. 14 vaguely reflects the 1/1 mode dominance range in the regime map, which agrees with the identification through wake visualisations, force evolutions and FFT spectra.

Notably, the hydrodynamic forces of 1/1 mode undergo complex variations with $\lambda^*$. For instance, $|C_{y.avg}|$ initially increases with $\lambda^*$ from 6.07 to 6.80, then diminishes till the end of 1/1 mode range. An opposite trend is shown in $C_{d2}K$ (Fig. 13$e$), where a local minimum of $C_{d2}K$ is found at the middle of 1/1 Arnold's tongue. The above variations align well with the transition of wake asymmetry as it horizontally passes through the 1/1 Arnold's tongue.

Fig. 13($e$) displays the impact of $A^*$ on $C_{d2}K$. With reducing the oscillation amplitude, the $C_{d2}K$ decreases for $\lambda^* < 5.0$ and becomes negative for small $A^*$, such as at $\lambda^* = 1.4 - 3.0$ and $A^* \lesssim 0.13$. As will be shown from the $C_{d2}K$ contour plot (Fig. 19$b$), the breadth of $\lambda^*$

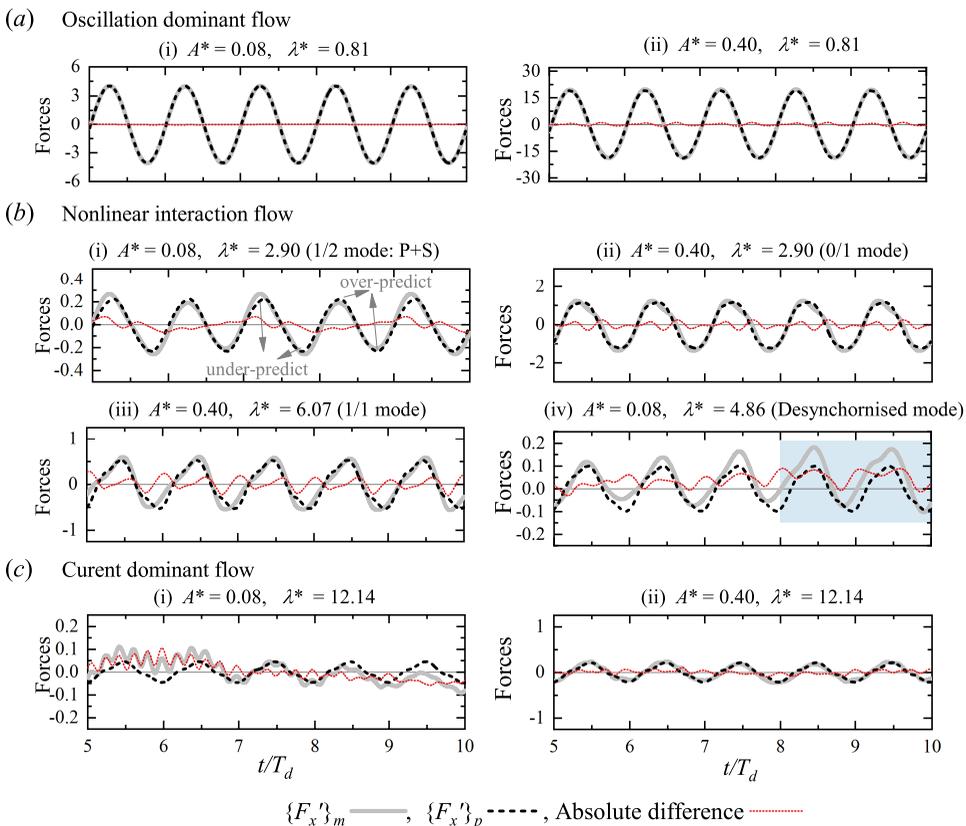

FIGURE 15. Evolution of fluctuating inline forces $F'_x$ for cases at $A^* = 0.08$ and $A^* = 0.40$. The $\{F'_x\}_m$ is the direct measurement of inline force in DNS; and the $\{F'_x\}_p$ is the reconstructed force using Eq. 2.4.

for $C_{d2}K < 0$ widens as $A^*$ decreases, covering an area of $\lambda^* = 1.2 - 3.5$ and $A^* < 0.20$. The collinear wake in this area can develop either synchronised modes, including 1/3, 1/2, 0/1, and desynchronised modes.

### 4.4. *Performance of the Morison-type formula*

The performance of Morison-type formula on estimating the inline force fluctuations is assessed firstly through the temporal evolutions of $F_x(t)$ from DNS (termed as 'measured force', $\{F_x\}_m$) with the reconstructed force from the independent model in Eq.2.4 (termed as 'predicted force', $\{F_x\}_p$). Since the time-averaged values of $\{F_x\}_m$ and $\{F_x\}_p$ are identical, following comparisons are focused on fluctuation components $F'_x(t)$. Fig. 15 showcases a few $F'_x(t)$ variations at $A^* = 0.08$ and $A^* = 0.40$. Each of the selected cases displays the measured $\{F'_x\}_m$ (solid grey line) and predicted $\{F'_x\}_p$ (dashed black line) over five successive oscillation periods, as well as their absolute difference (dotted red line).

The model performance varies greatly with $\lambda^*$. For oscillation dominant flows, such as cases in Fig. 15(a), the model gives excellent performance with negligible absolute difference compared to $F'_x$. In the nonlinear interaction group, the performance is affected by the wake formation. For instance, the P+S wake alters the $\{F'_x\}_m$ into a period-2 evolution (Fig. 15$b_i$). Since the Morison equation is proposed based on the period-1 assumption, the peaks of $\{F'_x\}_p$ in Fig. 15($b_i$) are slightly over-predicted in one period and under-predicted in the other period. In the case of 1/1 mode (Fig. 15$b_{iii}$), the poor prediction of $\{F'_x\}_p$ is because the $\{F'_x\}_m$ comprises frequency components other than $f/f_d = 1$, which cannot be considered in the Morison equation. Similarly, poor predictions are expected in desynchronised mode when the $\{F'_x\}_m$ exhibits aperiodic evolution (see blue shaded area in Fig. 15$b_{iv}$). For the current dominant flow (Fig. 15$c$), the performance is not acceptable for small $A^*$, but is getting better for higher $A^*$ due to the larger contribution of cylinder motion to $F_x$ (due to larger $U_m$).

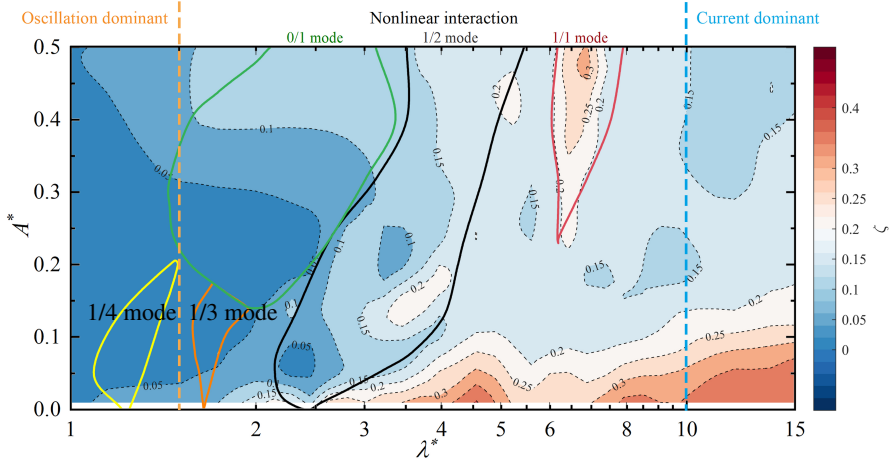

FIGURE 16. Contour of $\zeta$ over the $\lambda^* - A^*$ plane.

The root-mean-square difference $\zeta$ between $\{F_x\}_p$ and $\{F_x\}_m$ is calculated to give quantitative assessment, which is defined as,

$$\zeta = \frac{2}{\{F_x\}_m^{max} - \{F_x\}_m^{min}} \sqrt{\frac{\sum_{i=1}^{N}(\{F_x\}_m - \{F_x\}_p)^2}{N}}, \qquad (4.1)$$

where $N$ is the total sampling points over the statistical period. $\{F_x\}_m^{max}$ and $\{F_x\}_m^{min}$ are the maximum and minimum peaks of $\{F_x\}_m$. A lower value of $\zeta$ indicates a better prediction.

Fig. 16 depicts the $\zeta$ contour plot across the $\lambda^* - A^*$ space. Generally, $\zeta$ increases with higher $\lambda^*$. For large $\lambda^* > 10.0$ when the wake resembles the Kármán vortex street, the performance is generally worse with $\zeta$ reduces from about 0.4 to 0.15 as $A^*$ increases. In the intermediate range of $\lambda^* = 1.5 - 10$, $\zeta$ ranges around 0.05 - 0.3, displaying non-monotonic variations with both $\lambda^*$ and $A^*$. These complex distributions are attributed to the occurrence of different local wake characteristics in this flow group, such as discussions offered before. At $\lambda^* \lesssim 1.5$, all $\zeta$ values are below 0.1, indicating that the Morison equation predicts the inline force fluctuations with less than 10% root-mean-square difference. Particularly, the Morison equation performs exceptionally well for $A^* \approx 0.05 - 0.30$, with $\zeta$ less than 5%. The relatively poorer performance for $A^* \lesssim 0.05$ is because the total inline force at small $\lambda^*$ is mainly governed by the steady current, where the force induced by the cylinder oscillation constitutes a minor component of $F_x$ due to the lower $U_m$ compared to $U_\infty$. Consequently, $F_x(t)$ exhibits aperiodic variations within each oscillation cycle, leading to poor performance. The large value of $\zeta$ for $A^* > 0.30$ mirrors the behaviour of the 3-term MOJS for an oscillation-only case (see Fig. 10 in Ren *et al.* 2021), and is attributed to factors such as the development of wake asymmetry, additional vortex shedding, and the propagation of shed vortices.

Though the selection of acceptable estimation error $\zeta$ in engineering practice depends on the specific requirements for offshore structure design, Fig. 16 provides an indication to engineers of whether the use of Morison-type formula in the independent velocity model is capable of representing the inline force evolution using the data set of $[C_{x.avg}, C_{d1}, C_{d2}, C_m]$ predetermined from modelled cases. Considering a 10% root-mean-square difference, Fig.

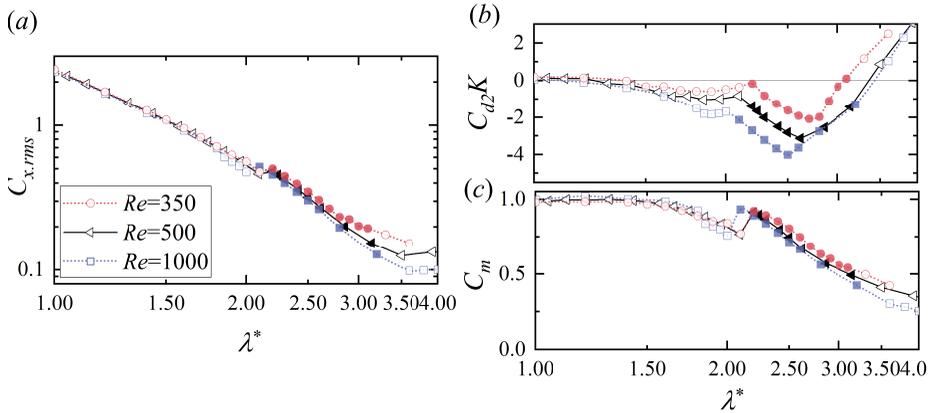

FIGURE 17. Evolution of (a) $C_{x.rms}$, (b) $C_{d2}K$ and (c) $C_m$ with $\lambda^*$ for $A^* = 0.05$ at $Re = 350 - 1000$. Cases with 1/2 mode are filled as solid symbols.

16 suggests the model is able to estimate the inline force at $\lambda^* \lesssim 1.5$, regardless of $A^*$. Outside of this range, the use of Morison-type formula requires cautions.

## 5. Discussions

The impact of $Re$ on collinear flow hydrodynamics is discussed here. In terms of the wake regime distribution, the present regime map in Fig. 2 at $Re = 500$ reveals consistent occurrence sequence for synchronised modes with those at low-$Re$ studies (Tang *et al.* 2017; Kim & Choi 2019). The Arnold's tongues of 0/1, 1/2 and 1/1 appear accordingly as $\lambda^*$ increases and occupy in similar $\lambda^* - A^*$ spaces. A noticeable difference of the present regime map with those at low $Re$, such as Fig. 15 in Kim & Choi (2019), is on the development of desynchronised modes between the 1/1 and 1/2 Arnold's tongues. Similar to the discussion related to Fig. 12, we attribute this discrepancy to the formation of 3-D wakes at high $Re$, resulting in irregular force evolutions.

To further illustrate the influence of $Re$ on hydrodynamic force coefficients and the performance of Morison equation, additional 3-D simulations are conducted at a fixed $A^* = 0.05$ and across a range of $Re = 350 - 1000$. Fig. 17 depicts the variations of $C_{x.rms}$, $C_{d2}K$ and $C_m$ with $\lambda^*$ for different $Re$. Across various $Re$ values, the occurrence of the 1/2 mode (filled symbols) is consistently observed at similar $\lambda^* \approx 2.10 - 3.20$. Higher Reynolds number slightly reduces the magnitudes of $C_{x.rms}$ due to the reduced contribution of viscous force to the total inline force. Similar decreasing trends are observed for $C_{d2}K$ and $C_m$ in Fig. 17 (b) and (c), respectively. The Morison equation works well for $\lambda^* \lesssim 1.50$ till $Re = 1000$, where $\zeta$ is less than 5% (not shown).

Above results suggest that general conclusions derived from collinear flows at $Re = 500$ remain valid across a broader range of Reynolds numbers up to 1000. However, extending the present analysis to higher Reynolds numbers would pose computational challenges due to increased computational costs with increasing $Re$. Further investigation is recommended to gain a deeper understanding of how the collinear flow hydrodynamics evolves with $Re$ for $Re > 1000$.

## 6. Conclusions

This study utilises three-dimensional (3-D) direct numerical simulations (DNS) to explore the wake and force characteristics of an oscillating cylinder in inline steady currents (collinear flow) over a wide parameter space of normalised amplitude $A^*$ from 0.01 to 0.50 and normalised wavelength $\lambda^*$ from 0.4 to 25.0. The Reynolds number is fixed at 500, aiming to extend the existing knowledge of collinear flow hydrodynamics to higher *Re* where 3-D wake develops.

Three distinct groups are identified based on their resemblance and distinction to two canonical scenarios: the steady current past a fixed cylinder (referred to as current-only) and the oscillating cylinder in quiescent fluid (referred to as oscillation-only). They include:

• Oscillation dominant flow ($\lambda^* \lesssim 1.5$): The steady current minimally impacts the wake induced by the cylinder oscillation. The flow along the cylinder surface is characterised by the symmetric generation of fine-scale and chaotic vortices by the oscillating cylinder and the inline force fluctuations resemble the oscillation-only counterparts. The steady current carries these vortices downstream of the cylinder, forming two parallel shear layers, before they interact to result in shedding of large coherent vortex structures. During this process, the wake three-dimensionality is significantly enhanced.

• Nonlinear interaction flow ($1.5 \lesssim \lambda^* \lesssim 10.0$): Complex interactions between the wake induced by the current and the oscillating cylinder lead to the formation of multiple synchronised modes, interleaved with non-synchronised modes. Distinct wake characteristics are observed across the $\lambda^* - A^*$ space, influencing hydrodynamic force coefficients to form non-monotonic variations. In total, five major Arnold's tongues related to the formations of $p/q$ synchronised modes, featured by $q$ cycles of vortex shedding on $q$ cycles of cylinder oscillation, are mapped in this space. The vortex shedding formations in each Arnold's tongue show strong dependency with both $\lambda^*$ and $A^*$.

• Current dominant flow ($\lambda^* \gtrsim 1.0$). The cylinder oscillation perturbs the conventional Kármán vortex street, resulting in a narrower and shorter wake than the current-only counterpart. Apart from RMS inline force coefficient, other hydrodynamic force coefficients close to the current-only counterparts.

The high-resolution hydrodynamic force contour plots on the plane of $\lambda^* - A^*$ are provided, revealing non-monotonic variations of hydrodynamic force coefficients. These variations are elucidated by linking them with dynamics of the wake.

The performance of independent force-velocity model, incorporating the 3-term Morison equation proposed by Ren *et al.* (2021), in predicting inline force fluctuations is assessed. For $\lambda^* \lesssim 1.50$ across all examined $A^*$ values, this model yields accurate estimations, with less than a 10% difference compared to the measured force from 3-D DNS. However, its performance diminishes beyond this range due to increased interactions of the steady current with wake characteristics, such as the occurrence of synchronised and desynchronized modes, distinct vortex shedding formations, and the development of 3-D structures.

The potential impact of Reynolds numbers on collinear flow hydrodynamics has been evaluated over a range of *Re* from 350 to 1000, affirming the validity of the general conclusions drawn from this study. The discovered collinear flow hydrodynamics could potentially provide insights to other fluid-structure interaction problems.

**Acknowledgements**

This work was supported by the Australia Research Council Discovery Grant (Project ID: DP200102804). This research was supported by computational resources provided by the National Computational Merit Allocation Scheme (NCMAS) and the Pawsey Supercomputing Centre with funding from the Australian Government and the Government of Western Australia.

**Appendix**

Fig. 18(*a-c*) display the contours of $C_{x.avg}$, $C_{x.rms}$ and $C_{y.rms}$, respectively. Contours of decomposed coefficients ($C_{d1}K$, $C_{d2}K$, and $C_m$) from Eq.2.4 are shown in Fig. 19.

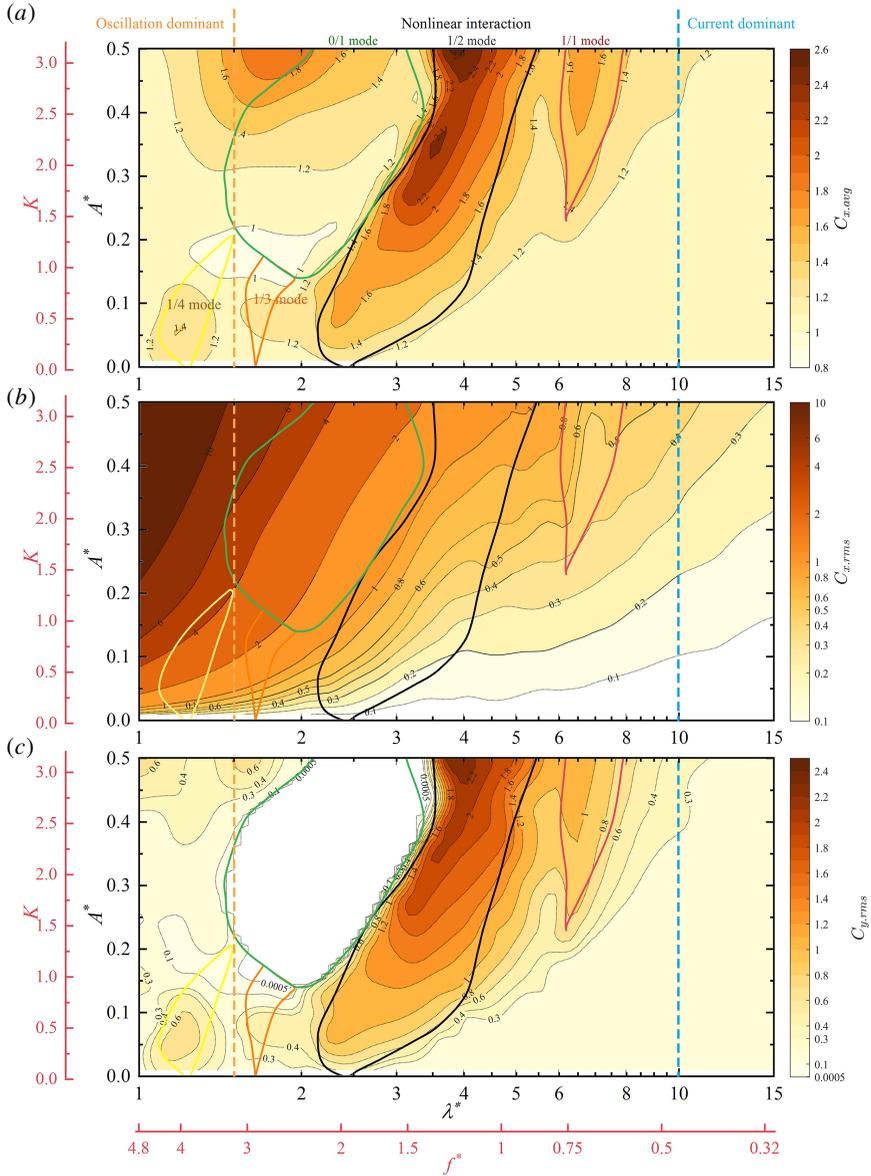

FIGURE 18. Contours of the (*a*) time-averaged inline force coefficient $C_{x.avg}$, (*b*) RMS inline force coefficient $C_{x.rms}$ and (*c*) RMS transverse force coefficient $C_{y.rms}$.

***Declaration of Interests.** The authors report no conflict of interest.*

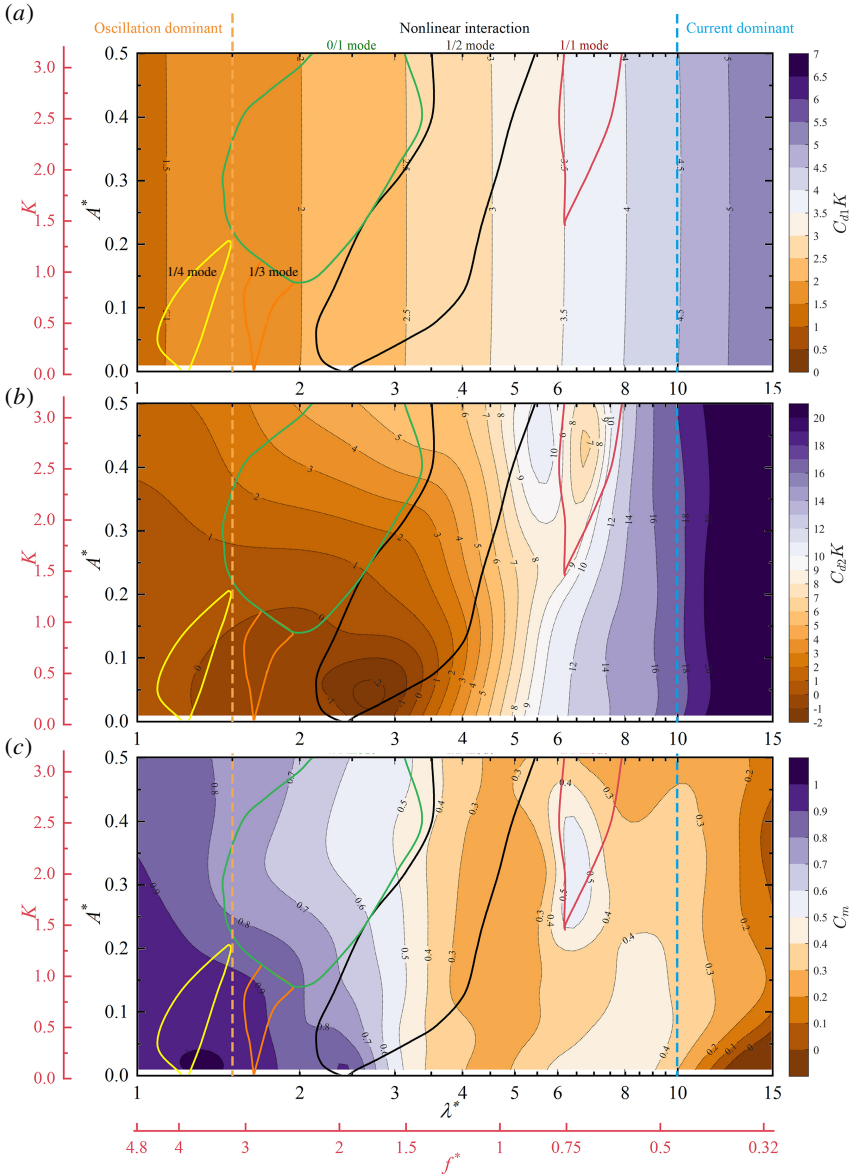

FIGURE 19. Contours of the (a) linear drag force factor $C_{d1}K$, (b) form drag force factor $C_{d2}K$ and (c) inertia force coefficient $C_m$ from Morison type formula in Eq. 2.4.